\documentclass[fleqn,twoside]{article}
\usepackage{espcrc2}

\usepackage{graphicx}

\newcommand{\AmS}{{\protect\the\textfont2
  A\kern-.1667em\lower.5ex\hbox{M}\kern-.125emS}}

\begin{document}
\title{BLACK HOLE REMNANTS AND DARK MATTER}

\author{Pisin Chen\address{Stanford Linear Accelerator Center \\
Stanford University, Stanford, CA 94309}\thanks{Work supported by
Department of Energy contract DE--AC03--76SF00515}
        and
        Ronald J. Adler\address{Gravity Probe B, W. W. Hansen Experimental Physics Laboratory\\
Stanford University, Stanford CA 94035}\thanks{Work partially
funded by NASA, Grant NAS8-39225}}

\begin{abstract}
We argue that, when the gravity effect is included, the
generalized uncertainty principle (GUP) may prevent black holes
from total evaporation in a similar way that the standard
uncertainty principle prevents the hydrogen atom from total
collapse. Specifically we invoke the GUP to obtain a modified
Hawking temperature, which indicates that there should exist
non-radiating remnants (BHR) of about Planck mass. BHRs are an
attractive candidate for cold dark matter. We investigate an
alternative cosmology in which primordial BHRs are the primary
source of dark matter. \vspace{1pc}
\end{abstract}

\maketitle

\section{INTRODUCTION}

In the standard view of black hole thermodynamics, based on the
entropy expression of Bekenstein\cite{bekenstein1} and the
temperature expression of Hawking\cite{hawking}, a small black
hole should emit black body radiation, thereby becoming lighter
and hotter, leading to an explosive end when the mass approaches
zero. However Hawking's calculation assumes a classical background
metric and ignores the radiation reaction, assumptions which must
break down as the black hole becomes very small and light. Thus it
does not provide an answer as to whether a small black hole should
evaporate entirely, or leave something else behind, which we refer
to as a black hole remnant (BHR).

Numerous calculations of black hole radiation properties have been
made from different points of view\cite{wilczek}, and some hint at
the existence of remnants, but in the absence of a well-defined
quantum gravity theory none appears to give a definitive answer.

A cogent argument against the existence of BHRs can be
made\cite{susskind}: since there is no evident symmetry or quantum
number preventing it, a black hole should radiate entirely away to
photons and other ordinary stable particles and vacuum, just like
any unstable quantum system.

In a recent paper\cite{acs}, we invoked the generalized
uncertainty principle (GUP)\cite{veneziano,adler,maggiore} and
argued the contrary, that the total collapse of a black hole may
be prevented by dynamics and not by symmetry, just like the
prevention of hydrogen atom from collapse by the uncertainty
principle\cite{shankar}. Our arguments then lead to a modified
black hole entropy and temperature, and as a consequence the
existence of a BHR at around the Planck mass. Here we first repeat
these arguments and derivations. We then investigate a cosmology
in which primordial BHRs serve as the primary source for dark
matter.

\section{GENERALIZED UNCERTAINTY \\ PRINCIPLE}

The uncertainty principle argument for the stability of hydrogen
atom can be stated very briefly. The energy of the electron is
$p^2/2m - e^2/r$, so the classical minimum energy is very large
and negative, corresponding to the configuration $p=r=0$, which is
not compatible with the uncertainty principle. If we impose as a
minimum condition that $p\approx \hbar/r$, we see that
$E=\hbar^2/2mr^2-e^2/r$, thus we find
\begin{equation}
r_{min}=\frac{\hbar^2}{me^2}\ , \quad {\rm and}\quad
E_{min}=-\frac{me^4}{2\hbar^2}\ . \label{eq:A}
\end{equation}
That is the energy has a minimum, the correct Rydberg energy, when
$r$ is the Bohr radius, so the atom is stablized by the
uncertainty principle.

As a result of string theory\cite{veneziano} or more general
considerations of quantum mechanics and
gravity\cite{adler,maggiore}, the GUP gives the position
uncertainty as
\begin{equation}
\Delta x\geq \frac{\hbar}{\Delta p}+ L_p^2\frac{\Delta p}{\hbar}
\quad\quad \Big(L_p=\sqrt{\frac{G\hbar}{c^3}}\Big)\ . \label{eq:K}
\end{equation}
A heuristic derivation may also be made on dimensional grounds. We
think of a particle such as an electron being observed by means of
a photon of momentum $p$. The usual Heisenberg argument leads to
an electron position uncertainty given by the first term in
Eq.(\ref{eq:K}). But we should add to this a term due to the
gravitational interaction of the electron with the photon, and
that term must be proportional to $G$ times the photon energy, or
$Gpc$. Since the electron momentum uncertainty $\Delta p$ will be
of order of $p$, we see that on dimensional grounds the extra term
must be of order $G\Delta p/c^3$, as given in Eq.(\ref{eq:K}).
Note that there is no $\hbar$ in the extra term when expressed in
this way. The position uncertainty has a minimum value of $\Delta
x=2L_p$, so the Planck distance plays the role of a fundamental
distance.

\section{STANDARD HAWKING EFFECT}

The Hawking temperature for a spherically symmetric black hole may
be obtained in a heuristic way with the use of the standard
uncertainty principle and general properties of black holes. We
picture the quantum vacuum as a fluctuating sea of virtual
particles; the virtual particles cannot normally be observed
without violating energy conservation. However near the surface of
a black hole there is an effective potential energy that is strong
enough to negate the rest energy of a particle and give it zero
total energy; of course the surface itself is a one-way membrane
which can swallow particles so that they are henceforth not
observable from outside. The net effect is that for a pair of
photons one photon may be absorbed by the black hole with
effective negative energy $-E$, and the other may be emitted to
asymptotic distances with positive energy $+E$.

The characteristic energy $E$ of the emitted photons may be
estimated from the uncertainty principle. In the vicinity of the
black hole surface there is an intrinsic uncertainty in the
position of any particle of about the Schwarzschild radius,
$\Delta x=r_s$, due to the behavior of its field
lines\cite{adler2} - as well as on dimensional grounds. This leads
to a momentum uncertainty
\begin{equation}
\Delta p \approx \frac{\hbar}{\Delta
x}=\frac{\hbar}{2r_s}=\frac{\hbar c^2}{4GM}\ , \label{eq:C}
\end{equation}
and hence to an energy uncertainty of $\Delta pc=\hbar c^3/4GM$.
We identify this as the characteristic energy of the emitted
photon, and thus as a characteristic temperature; it agrees with
the Hawking temperature up to a factor $2\pi$, which we will
henceforth include as a "calibration factor" and write (with
$k_B=1$),
\begin{equation}
T_{\rm H} \approx \frac{\hbar c^3}{8\pi GM}=\frac{M_p^2 c^2}{8\pi
M} \quad\quad \Big(M_p=\sqrt{\frac{\hbar c}{G}}\Big)\ .
\label{eq:E}
\end{equation}
We know of no way to show heuristically that the emitted photons
should have a thermal black body spectrum except on the basis of
thermodynamic consistency.

If the energy loss is dominated by photons we may use the
Stefan-Boltzmann law to estimate the mass and energy output as
functions of time. With use of the Hawking temperature and a mass
in units of the Planck mass, $x=M/M_p$, the rate of energy loss is
\begin{equation}
\frac{dx}{dt} = -\frac{1}{60(16)^2\pi T_p}\frac{1}{x^2}=
-\frac{1}{t_{ch}}\frac{1}{x^2}\ , \label{eq:I}
\end{equation}
where $T_p=(\hbar G/c^5)^{1/2}$ is the Planck time and
$t_{ch}=60(16)^2\pi T_p\approx 4.8\times 10^4 T_p$ is a
characteristic time for BH evaporation. It follows that the mass
and the energy output rate are given by
\begin{equation}
x(t)=\Big[x_i^3-\frac{3t}{t_{ch}}\Big]^{1/3}\ ,
\end{equation}
\begin{equation}
\frac{dx}{dt}=\frac{1}{t_{ch}(x_i^3-3t/t_{ch})^{2/3}}\ ,
\label{eq:J}
\end{equation}
where $x_i$ refers to the initial mass of the hole. The black hole
thus evaporates to zero mass in a time given by
$t/t_{ch}=(M_i/M_p)^3/3$, and the rate of radiation has an
infinite spike at the end of the process.

\section{BLACK HOLE REMNANTS}

We may use the GUP to derive a black hole temperature exactly as
in the previous section. This gives
\begin{equation}
\frac{\Delta p}{\hbar}=\frac{r_s}{2L_p^2}\Big[1\mp
\sqrt{1-2L_p^2/\Delta x^2}\Big]\ , \label{eq:L}
\end{equation}
and therefore
\begin{equation}
T_{\rm GUP} = \frac{Mc^2}{4\pi}\Big[1\mp\sqrt{1-M_p^2/M^2}\Big]\ .
\label{eq:M}
\end{equation}
This agrees with the Hawking result for large mass if the
negative sign is chosen, whereas the positive sign has no evident
physical meaning. Note that the temperature becomes complex and
unphysical for mass less than the Planck mass and Schwarzschild
radius less than $2L_p$, the minimum size allowed by the GUP. At
the Planck mass the slope is infinite, which corresponds to zero
heat capacity of the black hole.


The BH evaporation rate is
\begin{equation}
\frac{dx}{dt}=-\frac{16x^6}{t_{ch}}\Big[1-\sqrt{1-\frac{1}{x^2}}
\Big]^4 \ . \label{eq:O}
\end{equation}
Thus the hole evaporates to a Planck mass remnant in a time given
by
\begin{eqnarray}
\frac{t}{t_{ch}}=&-&\frac{1}{16}\Big[\frac{8}{3}x_i^3-8x_i-\frac{1}
{x_i}+\frac{8}{3}(x_i^2-1)^{3/2} \cr
&-&4\sqrt{x_i^2-1}+4\cos^{-1}\frac{1}{x_i} +\frac{19}{3}\Big]\ .
\label{eq:Q}
\end{eqnarray}
The energy output given by Eq.(\ref{eq:O}) is finite at the end
point when $x=1$ and is given by $dx/dt=-16/t_{ch}$, whereas for
the Hawking case it is infinite at the endpoint when $x=0$. The
present results thus appear to be more physically reasonable.


\section{BHRs AS DARK MATTER}

Black hole remnants (BHRs) are a natural candidate for cold dark
matter\cite{macgibbon} since they are a form of weakly interacting
massive particles (WIMPs)\cite{DM}.

The possible source and abundance of BHRs are of interest. The
most natural source is in primordial geometric fluctuations, which
would be sufficiently large only in the Planck era, at about the
Planck temperature. Rigorous derivations\cite{Gross} as well as
simple thermodynamic arguments\cite{Kapusta} imply that random
fluctuations would produce a Boltzmann distribution of black
holes, down to Planck mass, with a number density of $\sim
1/L_p^3$. In one version of standard inflationary
cosmology\cite{ohanion} the scale function increases by a factor
of about $10^{74}$ from the Planck era to the present, and since
the number density of matter scales as the cube of this we obtain
a number density that is down by about $10^{223}$, or
$10^{-118}/{\rm m}^3$. In comparison, the large scale density of
dark matter is roughly equal to the critical density, about
$2\times 10^{-26} {\rm kg/m}^3$, which implies a BHR number
density $\sim 10^{-18}/{\rm m}^3$. These are evidently
incompatible.

We have made some preliminary considerations of an alternative
cosmology containing primordial black holes. In the spirit of
Hartle and Hawking\cite{hartle} we suppose the universe was
initially a truly chaotic quantum foam system without ordinary
spacetime, and in particular without a time direction in that the
signature was (1,1,1,1). A fluctuation in the signature to (-1,
1,1,1) would then produce a time direction and turn it into an
exponentially expanding de Sitter space with heavy vacuum density
probably equal to the Planck density. At the very beginning of
time a thermal distribution of black holes would be
produced\cite{Gross}, and during the expansion would decay to
Planck size remnants and radiation. The presence of the black
holes and radiation (photons, gravitons, etc.) would change the
equation of state from that of heavy vacuum with $p=-\rho$, to a
mixture of BHR matter with $p=0$ and radiation with $p=\rho/3$,
and presumably very very little residual vacuum energy. This would
change the scale function from an exponential to a power-law, and
apparently not involve a horizon paradox. If the transition
involves a continuous energy density change, the scale function
would also be continuous and have a continuous derivative, so it
would change according to
\begin{equation}
a(t)=e^{t/T_p}\\ \to \\
a(t)=\Big(\frac{e}{n}\Big)^n\frac{t^n}{T_p^n}\ \,
\end{equation}
at $t=nT_p$, where $n$ should be between $1/2$ for radiation and
$2/3$ for matter. The duration of exponential expansion would thus
be quite short. Very roughly the decrease in number density of
BHRs from the beginning of time, $t=0$, to the present time,
$t=t_0$, would then be
\begin{equation}
\Big[\frac{a(t_0)}{a(0)}\Big]^3\approx
\Big(\frac{t_0}{T_p}\Big)^{3n} \ .
\end{equation}
If $n$ were about $2/3$, appropriate to matter, then the scale
function would decrease by about $10^{41}$ and the density factor
by about $10^{123}$, which implies a present number density of
remnants of
\begin{equation}
\rho_{\rm BHR}\approx 10^{-18}{\rm m}^{-3} \ ,
\end{equation}
which is the value needed. However if a radiation dominated period
of expansion is included, as it apparently must, then $n$ should
be about 1/2 until the decoupling time $t_d$, and the present
density would be about $10^{10}/{\rm m}^3$, which is far too
large. We therefore need to include an ad hoc period of inflation
to obtain a reasonable density. Specifically, if we extend the
period of inflation from $nT_p$ to $\eta T_p$, followed by a
period of radiation dominance to $t_d$, and then matter dominance
to the present (but do not ask that the scale function have a
continuous derivative), we obtain roughly
\begin{equation}
\frac{a(t_0)}{a(0)}\approx
\frac{e^{\eta}}{\sqrt{\eta}}\Big(\frac{t_0^{2/3}}{T_p^{1/2}t_d^{1/6}}\Big)\ .
\end{equation}
This gives the desired density if the number of e-folding times is
chosen to be about $\sim 27$, roughly half the number usually used
in standard inflation. Thus this scenario shares features with the
standard inflation scenario, that the period of inflation and the
manner in which the vacuum energy is converted to radiation and
particles are ad hoc.

\section{CONCLUSION}

We have argued the logical existence of black hole remnants based
on the generalized uncertainty principle. We then investigated an
alternative cosmology in which BHRs are the primary source of dark
matter. The initial study indicates that our scenario is not
inconsistent with basic cosmological facts, but more scrutiny is
required before it can become a viable option.

%
%
%
%
%
%
%

\end{document}